# Experimental Evaluation of the Rheological Properties of Veriflex® Shape Memory Polymer

J. Klesa[1*], V. Placet[2], E. Foltête[2], M. Collet[2]
[1] Department of Aerospace Engineering, Czech Technical University in Prague, Czech Republic
[2] Department of Applied Mechanics, FEMTO-ST Institute, Besançon, France

**Abstract.** Shape memory polymers (SMPs) are materials with a great potential for future use in smart materials and structures. When heated from cold state (below the transformation temperature, which can either be the glass transition temperature or the melting temperature of the polymer) to hot state (above the transformation temperature) they undergo transformation which can be compared with martensitic transformation of shape memory alloys. This process induces great changes of the mechanical properties and some shape memory phenomenon can be observed. This study is an experimental evaluation of the mechanical properties of SMP Veriflex® under different test conditions. Veriflex® was chosen because of its easy accessibility. Furthermore its properties are similar to epoxy resins which make it very suitable for usage in a wide variety of technical applications. Dynamic mechanical analysis (DMA) was used to determine evolution of the viscoelastic properties versus temperature and frequency under cyclic harmonic loading. The glass transition temperature clearly appears in a range from 45°C to 60°C depending on loading frequency. The glass transition is noticeably marked by an impressive decrease in the storage modulus of about 4 decades. The master curve of Veriflex® was created and allows the time-temperature superposition to be constructed for this material. Thermo-mechanical working cycle of SMP with 100% elongation was also experimentally tested. Finally results from all these experimental investigations were used to design a demonstrator showing the possibility of application in engineering and especially for shape control.

## 1. Introduction

Shape memory polymers (SMPs) belong to the class of smart materials. They have the ability to memorize temporary shapes (intended or accidental deformation caused by an external force) and to recover their original and permanent shape upon an external stimulus, most typically thermal activation. These materials have found growing interest throughout the last few years considering their great potential for future. They surpass the metallic shape memory alloys (SMA) in their shape-memory properties. SMPs particularly offer deformation to a much higher degree, light weight, easy processability and cheapness compared to SMA [4]. They can be used in a large variety of applications, e.g. actuators, electromechanical systems, clothing manufacturing, morphing and deployable space applications, control of structures, self healing, biomedical devices and so on. The following review papers summarise current knowledge and recent progress relevant to the field of shape memory polymers: [1, 2, 4]. The shape memory mechanisms has been partly described in recent studies [4, 6] and is pretty well understood at the molecular level The knowledge and understanding of the physical properties and SMP's capabilities and limitations is required for an efficient employ in structural applications.

SMP Veriflex® manufactured by U.S. company CRG Industries was chosen for its easy supplying on the market, and can be bought in the form of resin and has reasonable price. For these reasons, this material can be easily used for structure design. The aim of this paper is to investigate the rheological properties of SMP Veriflex® and determine its glass transition temperature. Actually, a critical parameter for a shape memory polymer lies in its shape memory transition temperature [6]. Dynamic mechanical analysis (DMA) was used for experimental evaluation of rheological properties. Experimental demonstration of SMP working cycle is also presented. Information about functional fatigue of Veriflex® can be found in [5].

## 2. Rheological properties of Veriflex material

### 2.1 Material and method

A DMA Bose Electroforce 3200 was used to perform rheological experiments on Veriflex specimens (Fig. 1). This instrument employs a moving magnet linear motor to apply the solicitations to the sample.

[a] e-mail: klesa@aerospace.fsik.cvut.cz

The applied force is measured with a load sensor of 22 N with a resolution of about 10 mN, and the displacement is measured using a LVDT with a resolution of 1 μm. A small static strain is applied to the sample in order to maintain the specimen under a net tension during all the experiment. A sinusoidal force is applied around this mean value of strain. A software (Wintest) pilots the DMA and performs the viscoelastic properties calculation from force and displacement signals, and the sample size. A climatic chamber allows the temperature of the sample to be controlled with a temperature stability of about +/- 0.3°C. The temperature is measured by a thermocouple placed a few millimetres from the sample.

The cross section of specimen is 5*5 mm and its length 80 mm. At each time, the length value used to calculate the strain is the actual length of the sample and not the initial length. This "corrected sample length" allows the length variations due to thermal dilatation and viscoelastic creep to be corrected. Temperature varies between 30°C and 80°C, the frequency of the solicitation from 0.01 Hz to 10 Hz. Viscoelastic properties (storage modulus, loss modulus and loss factor) are measured every 2°C in isothermal conditions. The temperature is stabilised during 5 minutes before measurements. The heating rate is about 3°C per minute between each plateau temperature.

The clamping of the Veriflex® specimen is not really trivial because of the large variation of the samples stiffness and expansion according to temperature. To avoid any slippage in the specimen's attachment areas, specific clamps with compression springs were used. For the same reasons, it is not possible to control the actuator using a classical feedback loop (with displacement or load control). The test is directly controlled using the current of the actuator.

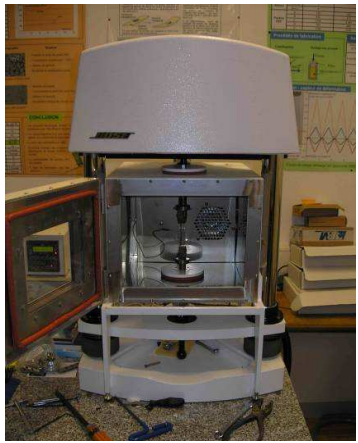
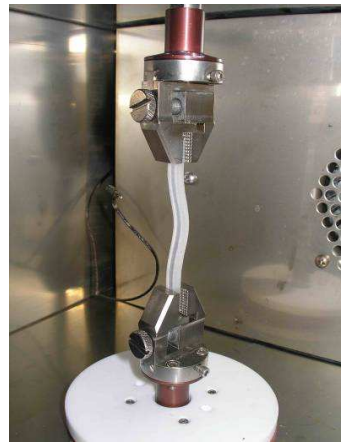

a)                                                    b)

**Fig. 1.** a) Dynamic Mechanical Analyser (DMA) used for experiments, b) Specimen clamping

## 2.2 Dependence of Veriflex® Viscoelastic Properties on Temperature

Results of dynamic mechanical analysis are commonly presented as storage modulus (E'), loss modulus (E'') and loss factor (tanδ = E''/E'). The storage modulus describes the capacity of material to support a load, and so represents the elastic part of the sample. The loss modulus is the viscous response of the sample and is proportional to the dissipated energy. The loss factor characterises the damping capacity of the material.

Figure 2 depicts the evolution of the viscoelastic properties according to temperature. Notice the huge decrease of storage modulus E' between 20°C and 90°C (more than 3 decades), a peak of E'' and a peak of tanδ. This temperature range at which a maximum of energy is dissipated denotes the largest movements of the polymer chains. This temperature is the glass transition temperature, noted Tg. It indicates that molecular movements have the same relaxation time as the applied frequency.

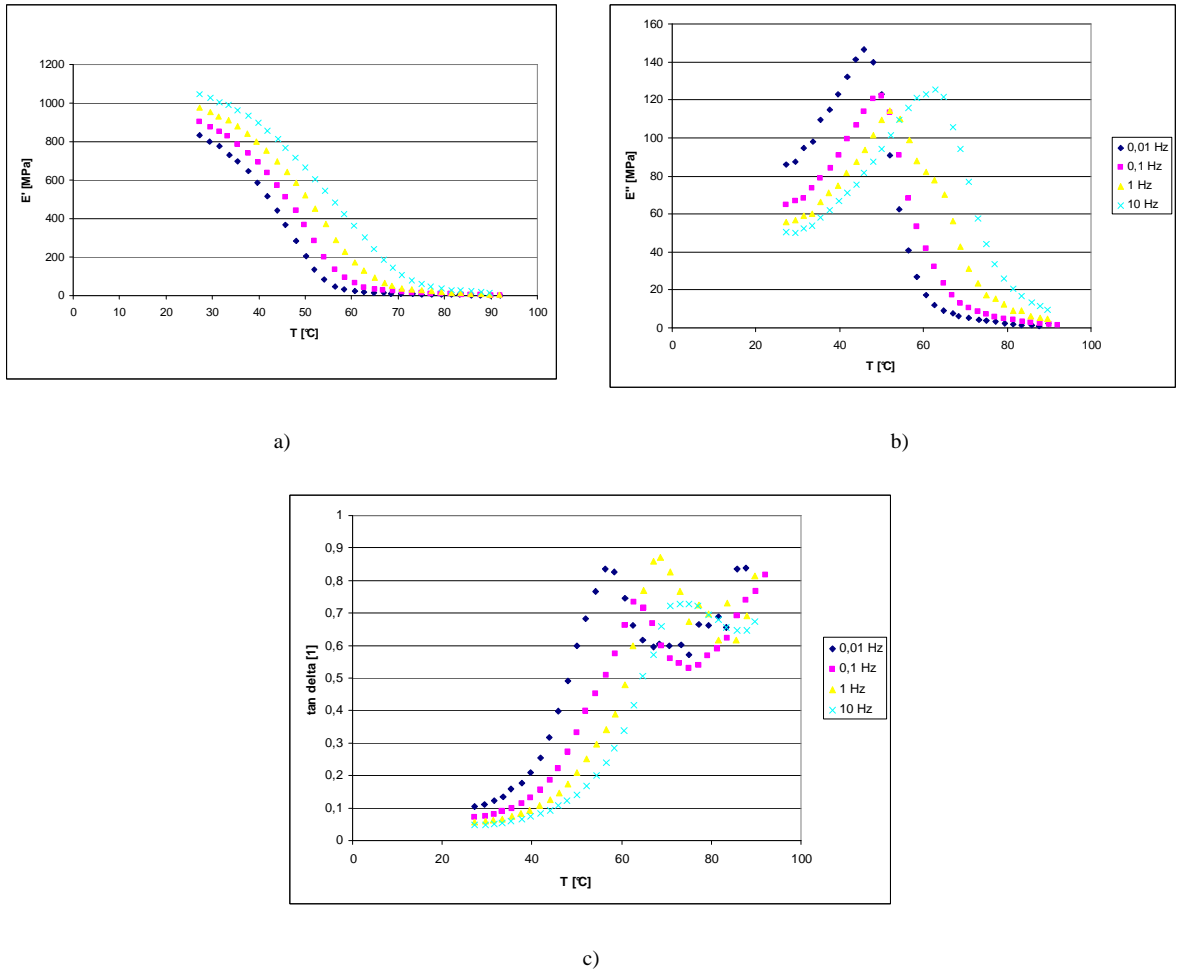

a)

b)

c)

**Fig. 2.** Viscoelastic properties of Veriflex samples according to temperature and frequency. a) Storage modulus (E') vs. temperature, b) Loss modulus (E'') vs. temperature, c) *tan δ* vs. temperature

$T_g$ was determined for the different frequencies using two different methods: the peak of E'' and the peak of tanδ. In the area of the peak, the curves were approximated by a polynomial curve and the peak was determined as the maximum of the polynomial regression on the studied temperature range. In some cases it was impossible to find the peak of tan δ and so only the peak of the loss modulus E'' was found. The results are summarised in tab. 1.

**Table 1.** Glass transition temperature of Veriflex® determined using two different methods (peak of loss modulus $E''$ and peak of tanδ), three samples were used, data in the table are mean value, minimum and maximum value of measured glass transition temperature for each frequency.

| $f$ [Hz] | 0.01 | 0.02 | 0.03 | 0.06 | 0.1 | 0.2 | 0.3 | 0.6 | 1 | 2 | 3 | 6 | 10 |
|---|---|---|---|---|---|---|---|---|---|---|---|---|---|
| $T_g$ from $E''$ [°C] | 45.5 | 46 | 47.5 | 48.5 | 49 | 50.5 | 51 | 52.5 | 54 | 55.5 | 58 | 59 | 61.5 |
| $T_g$ from tanδ [°C] | 58.5 | 61 | 61.5 | 62.5 | 64.5 | 65.5 | 67 | 68.5 | 70 | 71 | | | 75 |

As for all viscoelastic solid materials, the storage modulus of Veriflex® material increases with frequency. Moreover, the maximum value of the loss factor increases with frequency, which attests that the transition temperature shifts to higher values as the frequency increases. Graphic interpretation of the dependence of the glass transition temperature on frequency can be seen in fig. 3. The reproducibility of the measurements was checked using 3 samples.

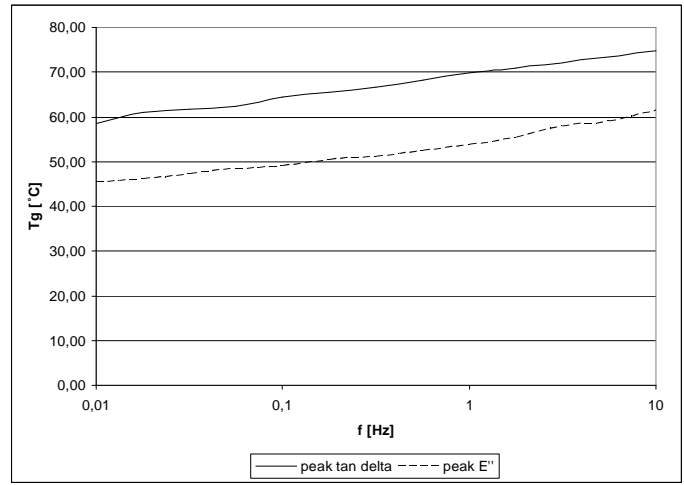

**Fig. 3.** Dependence of the glass transition temperature on the frequency of the loading (glass transition temperature was determined as the peak of the loss modulus E'' and at the same time as the peak of tanδ), see also table 1.

## 2.4 Master Curve

In polymer rheology, the time-temperature equivalence is an important concept. The thermal activation can be expressed by the WLF law (Williams, Landel, and Ferry)**.** This law is only valid over a certain range of temperature. Because the WLF law originates from the free volume theory, it is valid for temperature values above the softening temperature. Actually, molecular motion at the relaxation level needs volumetric expansion. The first step of the WLF analysis consists in building the master curve based on raw data. To do this, the curves of the viscoelastic properties (E', E'' and tanδ) vs. frequency at various temperatures are shifted along the x-axis to build up a single curve (Fig. 4). The shift value of each curve obtained at temperature T from the reference curve obtained at temperature $T_0$ defines the shift factor $a_{T,To}$.

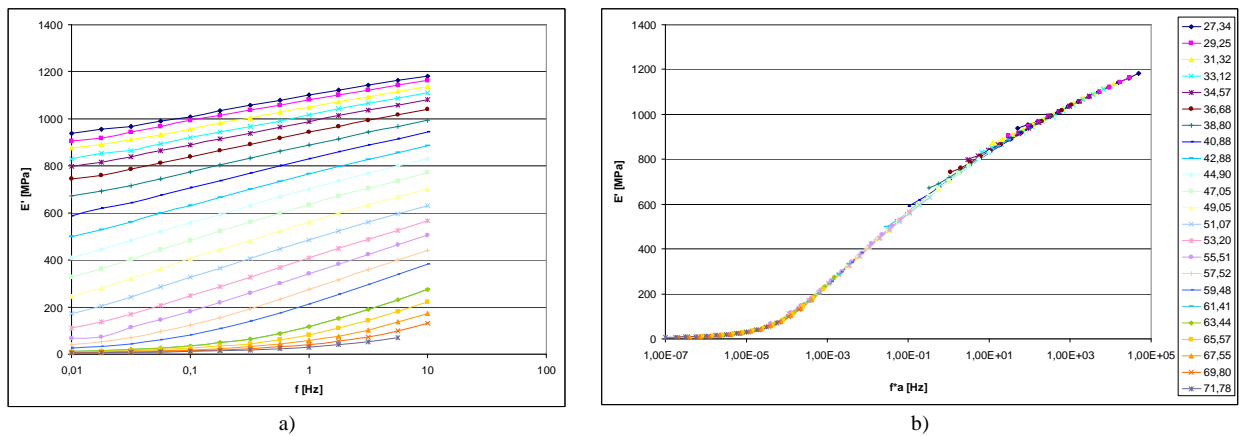

a)                                                                              b)

**Fig. 4.** a) Storage modulus (E') vs. frequency for different temperatures, b) Veriflex® master curve for storage modulus (E') (measurement multifrequency 3).

In Figure 5 master curves are depicted for all the viscoelastic properties, i.e. E', E'' and tanδ. The reference temperature for the presented master curves is 45˚ C. The dependence of the shift factor on temperature can be expressed as following according to Williams, Landel and Ferry:

$$\log a_{T,T_0} = \frac{-c_1^0 (T - T_0)}{c_2^0 + T - T_0}$$

(1)

where $a_{T,T_0}$ is the shift factor, $T_0$ the reference temperature (K), $c_1^0$ and $c_1^0$ two WLF constants. If the reference temperature is $T_g$, the two WLF constants are noted $C_1$ and $C_2$, and are supposed to be universal for each polymer.

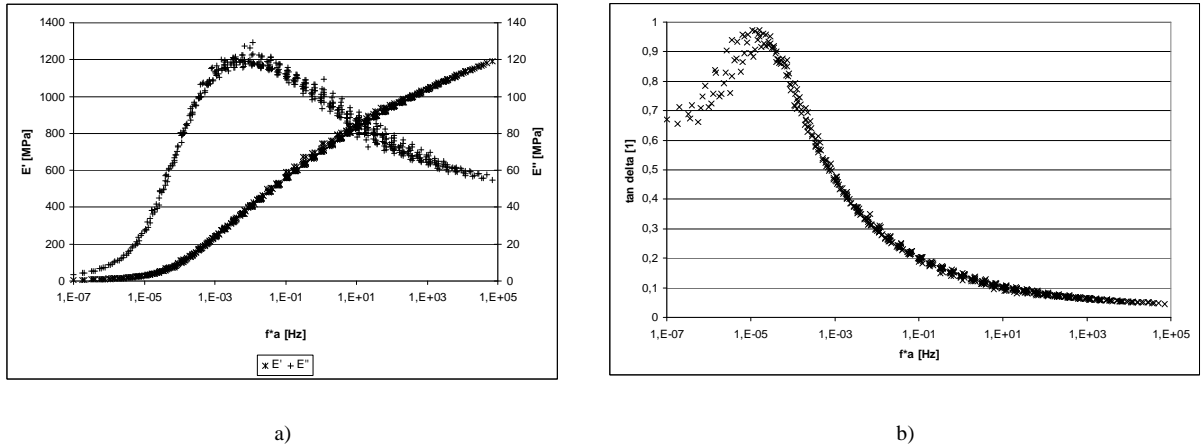

a)                                        b)

**Fig. 5.** a) Veriflex® master curves for storage modulus (E') and loss modulus (E''), b) Veriflex® master curve for tan$\delta$.

The WLF constants ($C_1$ and $C_2$) were determined from the curve $\log a_T$ versus $(T - T_0)$ (Fig. 6). The values of the WLF parameters are: $C_1$= 18.30, $C_2$ = 70.54 K and the activation energy $\Delta H_a$ = 502.1 kJ.mol$^{-1}$. The WLF parameters were determined for a temperature reference corresponding to the softening temperature of the verifmex polymer at very low frequency. The values of these constants are closed to the universal ones (for example, $C_1$= 14.5, $C_2$ = 50.4 K for polystyrene according to Oudet [3]). From the figure 6, one may consider the WLF law is valid between $T_g$ and $T_g$+45K. Below the softening temperature, the free volume is independent of temperature; the mobility variation with temperature can be determined following an Arrhenius law.

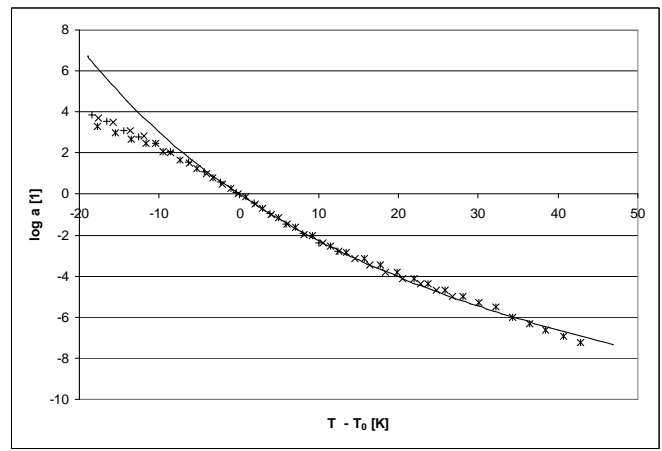

**Fig. 6.** Approximation of dependence of horizontal shift factor $a_{T,T_0}$ on temperature by WLF formula, $T_0$ = 45˚ C.

# 3. Veriflex® Working Cycle Demonstration

## 3.1 Experimental Setup

A universal commercial testing machine (Instron 6025) was used to carry out the demonstration of Veriflex® working cycle (fig. 7a). The machine is equipped with a thermal chamber to control the temperature of the sample during the test. The temperature is measured by a thermocouple placed in a piece of Veriflex® in the clamps near the specimen, which can be seen in upper parts of fig. 7b and 7c. Dog-bone shaped specimens (showing a reduced gauge section and enlarged shoulders) were produced from the polymer plate. The gauge length was 45 mm, and the rectangular cross section around 15 x 5 mm². The samples were clamped using wedge action grips. The samples were subjected to a displacement or a load control. The strain of sample was calculated from the crosshead displacement, which is measured by an optical encoder, with a resolution of 10 microns. The specimen in the testing machine can be seen in fig. 7b - without deformation and fig. 7c - deformed state. Reproducibility of the measurements was checked on several samples.

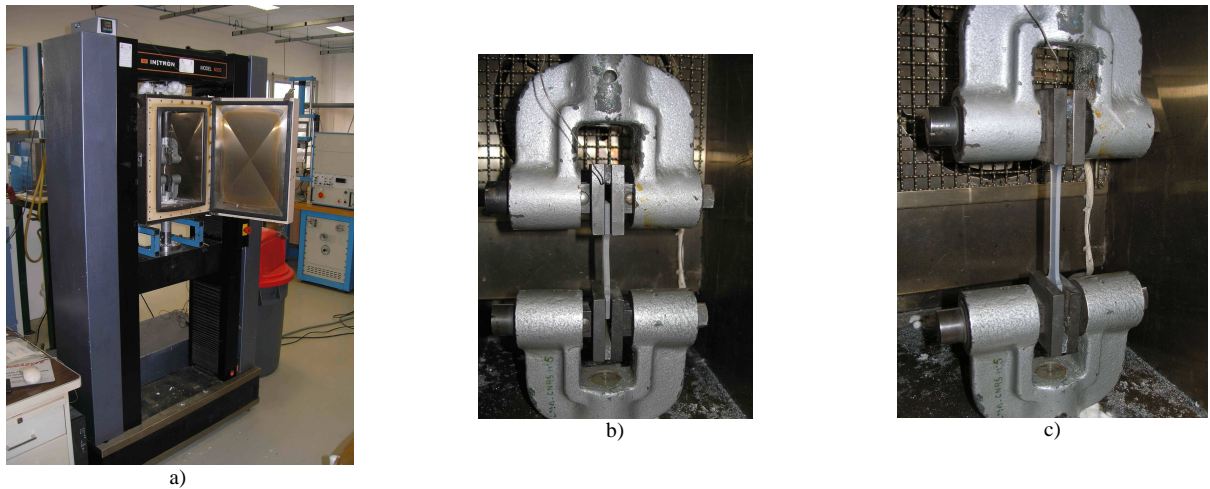

a)                                    b)                                    c)

**Fig. 7.** a) Instron testing machine used for Veriflex® working cycle demonstration, b) Specimen in the testing machine before the test, c) Specimen in the testing machine at the deformed state during the test.

### 3.2 Results

Experimentally tested working cycle of Veriflex® shape memory polymer is shown in Fig. 8. It consists of the following parts:

| | |
|---|---|
| 1 | Permanent shape, polymer is heated above transformation temperature, no deformation and no force is applied |
| 1 – 2 | Deformation from permanent shape 1 to temporary shape 2 at high temperature (= temperature higher that the transformation temperature of the polymer) |
| 2 – 3 | Relaxation |
| 3 – 4 | Cooling at constant applied force |
| 4 – 5 | Applied force is lowered to zero |
| 5 – 6 | Heating with zero force applied, polymer returns to its permanent shape |

Point 6 is close to point 1. Residual deformation (2 to 5 %) can be observed, but it is small compared with the large deformation at the point 2, which is approximately 100% (according to the Veriflex® datasheet, the maximal elongation is about 200%).

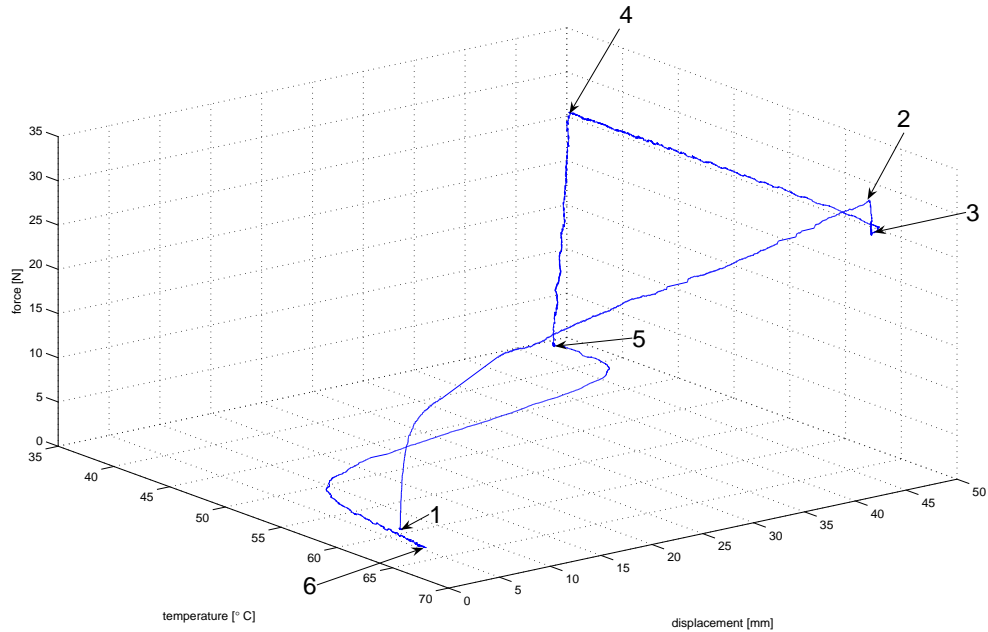

**Fig. 8.** Veriflex® working cycle in displacement, temperature and force coordinates.

# 4. Conclusion

Experimental results of DMA analysis on a Shape Memory Polymer are presented. The dependence of glass transition temperature on frequency of cyclic harmonic loading is obtained from these results. This provides much more detailed information than what can be found in the material datasheet. Time-temperature superposition was used to create the Veriflex® master curve. Working cycle of SMP Veriflex® was also experimentally demonstrated on a testing machine. The results of all these experiments will be used for designing a morphing structure demonstrator using SMP.